\documentclass{jetpl}
\twocolumn

\usepackage{amsmath,amssymb,epsfig,color,pstricks,bm,graphics,alltt}

\usepackage[colorlinks,plainpages=false,linkcolor=black,urlcolor=blue,citecolor=black,pdfpagemode=UseNone,pdfstartview=FitBH]{hyperref}

\definecolor{MyDarkBlue}{rgb}{0,  0.3,  0.9}
\definecolor{MyDarkBlack}{rgb}{0,  0,  0}

\begin{document}

\lat

\title{Electronic and Magnetic Properties of the New Iron -- Based 
Superconductor [Li$_{1-x}$Fe$_x$OH]FeSe}

\rtitle{Electronic and Magnetic Properties of New Superconductor [Li$_{1-x}$Fe$_x$OH]FeSe}

\sodtitle{Electronic and Magnetic Properties of the New Iron -- Based Superconductor [Li$_{1-x}$Fe$_x$OH]FeSe}

\author{$^a$I.\ A.\ Nekrasov\thanks{E-mail: nekrasov@iep.uran.ru}, $^{a,b}$M.\ V.\ Sadovskii\thanks{E-mail: sadovski@iep.uran.ru}}

\rauthor{I.\ A.\ Nekrasov, M.\ V.\ Sadovskii}

\sodauthor{I.\ A.\ Nekrasov, M.\ V.\ Sadovskii }

\sodauthor{I.\ A.\ Nekrasov, $M.\ V.\ Sadovskii}

\address{$^a$Institute for Electrophysics, Russian Academy of Sciences, 
Ural Branch, Amundsen str. 106,  Ekaterinburg, 620016, Russia\\
$^b$Institute for Metal Physics, Russian Academy of Sciences, Ural Branch,
S.Kovalevskoi str. 18, Ekaterinburg, 620290, Russia
}


\abstract{
We present the results of paramagnetic LDA band structure calculations: 
band dispersions, densities of states and Fermi surfaces, for the new iron 
based high-temperature superconductor LiOHFeSe.
Main structural motif providing bands in the vicinity
of the Fermi level is FeSe layer which is isostructural to the bulk FeSe 
prototype superconductor. The bands crossing the Fermi level and Fermi surfaces 
of the new compound are typical for other iron based superconductors.
Experimentally it was shown that introduction of Fe ions into LiOH layer
gives rise to ferromagnetic ordering of the Fe ions at $T_C$=10K. To study 
magnetic properties of [Li$_{0.8}$Fe$_{0.2}$OH]FeSe system we have performed 
LSDA calculations for $\sqrt 5 \times \sqrt 5$ superlattice and found 
ferromagnetism within the Li$_4$Fe(OH) layer. To estimate the Curie temperature 
we obtained Fe-Fe exchange interaction parameters for Heisenberg model from our 
LSDA calculations, leading to theoretical value of Curie temperature 10.4K in
close agreement with experiment.
}


\maketitle

\section{Introduction}

Novel iron based high-temperature superconductors 
\cite{kamihara_08} have been stimulating lots of experimental and theoretical 
studies since 2008 (extended reviews can be found 
in~\cite{UFN,Hoso_09,Johnson,FeSe,Mazin}). There are two large groups of 
iron -- based  superconductors: pnictides \cite{UFN,Hoso_09} and 
chalcogenides \cite{FeSe}. Electronic band structures of these 
superconductors, as well as some related systems, were compared in 
Refs. \cite{PvsC,relcomp}. 

\begin{figure}[!hb]
\center{
\includegraphics[clip=true,width=0.6\textwidth]{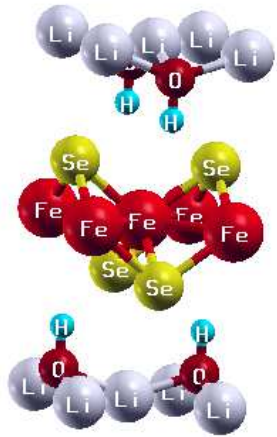}
}
\caption{Fig. 1. Crystal structure of LiOHFeSe.} 
\end{figure}

Among recently synthesized systems particularly interesting is 
[Li$_{1-x}$Fe$_x$OH](Fe$_{1-y}$Li$_y$)Se ($x\approx 0.2$, $y\approx 0.08$) 
compound with the temperature of superconducting transition $T_c$=43K and 
ferromagnetic ordering at Curie temperature $T_C$=10K \cite{Pachmayr}.

In this paper we present LDA calculated electronic structure, densities of 
states and Fermi surfaces for LiOHFeSe system in paramagnetic phase. 
To investigate ferromagnetism of Fe ions introduced into LiOH layer 
we performed LSDA calculations for the [Li$_{0.8}$Fe$_{0.2}$OH]FeSe 
compound. Corresponding values of Heisenberg model exchange parameters 
obtained from LSDA results were used to calculate Curie temperature, producing
a nice agreement with experiments.

\section{Electronic structure}

\begin{figure*}[!ht]
\includegraphics[clip=true,width=0.9\textwidth]{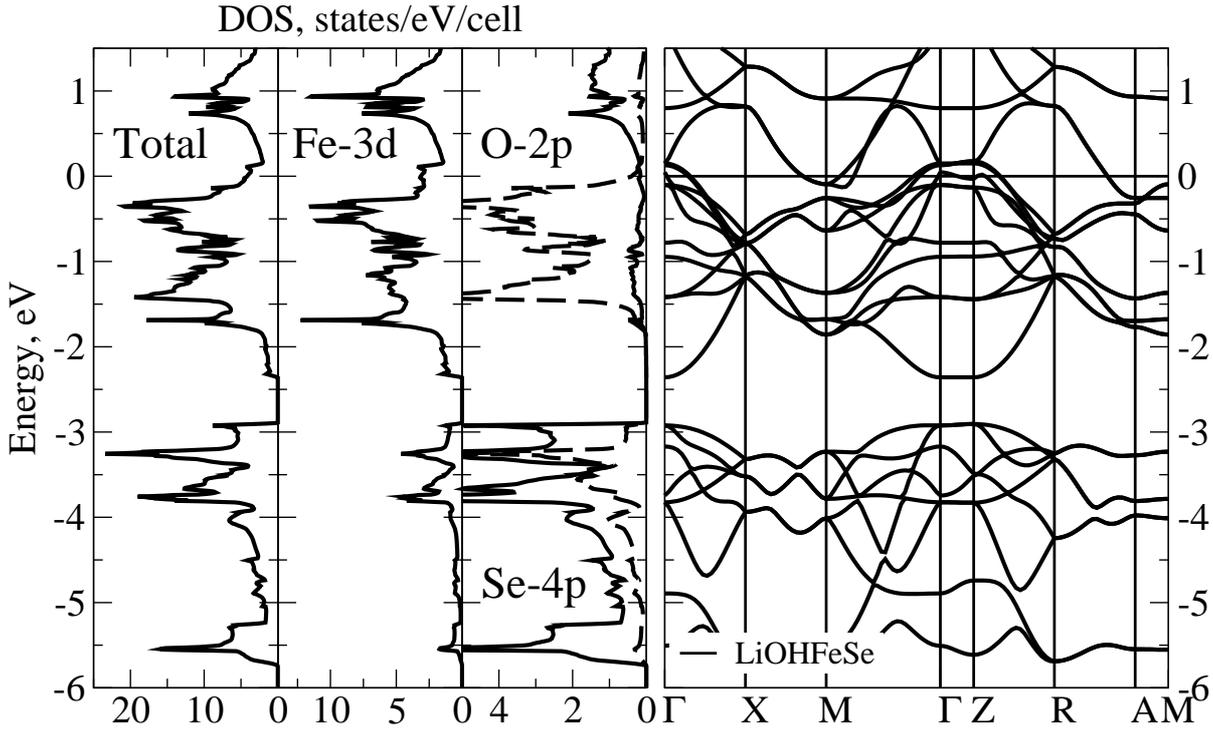}
\caption{Fig. 2. LDA calculated band dispersions (right) and densities of 
states (left) of paramagnetic LiOHFeSe. The Fermi level $E_F$ is at zero energy.} 
\end{figure*}

The crystal structure of [Li$_{1-x}$Fe$_x$OH](Fe$_{1-y}$Li$_y$)Se belongs to the
primitive tetragonal {\em P4/nmm} space group (see table S1 in supplementary 
materials of Ref. \cite{Pachmayr}). The LiOHFeSe system has quasi 
two--dimensional crystal structure of alternating LiOH and FeSe layers.
In Fig. 1 we present an idealised crystal structure of 
[Li$_{1-x}$Fe$_x$OH](Fe$_{1-y}$Li$_y$)Se. Here we neglect the presence of Li 
ions in FeSe layers as well as the presence of Fe ions in the LiOH layers.
Also we took Li ion in the 2a Wyckoff position instead of 4f one.
LiOH layer is formed by square lattice of Li ions with OH forming tetrahedra 
around them. FeSe layer consists of square lattice of Fe ions which are 
surrounded by tetrahedrally coordinated by Se ions. Actually, this
FeSe layer is isostructural to bulk FeSe material \cite{FeSe}.

For these idealised crystal structure we have performed LDA band structure 
calculations within the linearized muffin-tin orbitals method 
(LMTO)~\cite{LMTO} using default settings.

In Fig.~2 we show the calculated LDA band dispersions (on the right) and 
densities of states (DOS) (on the left). 
Electronic bands crossing the Fermi level are mostly formed by Fe-3d orbitals
and have bandshapes similar to previously studied BaFe$_2$As$_2$ and FeSe 
systems (see Refs. \cite{PvsC,Nekr2,SinghFeSe}).
Surprisingly O-2p states are quite close to the Fermi level, beginning
just below the Fermi level and going down in energy up to -2 eV.
However, there is almost no hybridization between O-2p and Fe-3d states.
The Se-4p orbitals form bands at energies below -3 eV.
Hybridization between Fe-3d and these Se-4p states is also rather small. 

\begin{figure}[!h]
\includegraphics[clip=true,width=0.45\textwidth]{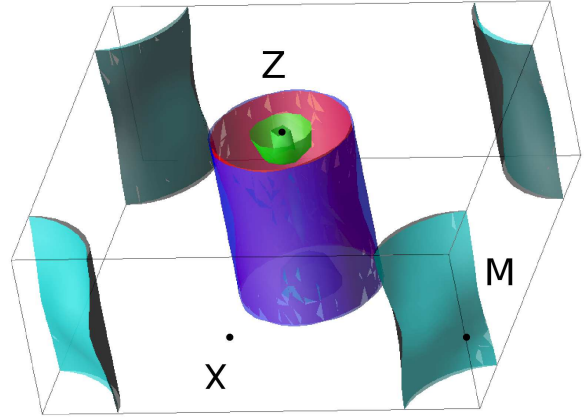}
\caption{Fig. 3. LDA calculated Fermi surfaces of LiOHFeSe within the first 
Brillouin zone.} 
\end{figure}

The value of LDA calculated total density of states of LiOHFeSe system at the 
Fermi level is $N(E_F)$=4.14 states/cell/eV. Performing simple BCS-like
estimates of superconducting critical temperature along the lines of 
Ref. \cite{PvsC} we obtain T$_c$=36K, which is somewhat lower
than the experimental value of $T_c$=43K  \cite{Pachmayr}.

In Fig.~3 LDA present the LDA calculated Fermi surfaces (FS) of LiOHFeSe.
In general a shape of the  FS of LiOHFeSe is typical to iron pnictides or 
chalcogenides \cite{PvsC,Nekr2,SinghFeSe}.
It has well established cylinders around  $\Gamma$ and $M$ points. Around
$\Gamma$ point there are two large almost degenerate ideal hole cylinders.
Electron cylinders around $M$-point are slightly corrugated. Also near 
the $\Gamma$-point there is a tiny hole pocket. Some rather small FS sheet 
appears also around the $Z$-point.

\section{Magnetic structure}

As reported in Ref. \cite{Pachmayr} the introduction of Fe ions into
LiOH layer leads to the appearance of ferromagnetism within the layer with 
Curie temperature about 10K. To mimic the experimental chemical composition 
with doping level $x$ about 0.2 we performed LSDA calculations for 
$\sqrt 5 \times \sqrt 5$ superlattice, analyzing the several possible 
magnetic structures, corresponding to different distributions of Fe ions in 
Li$_4$Fe layer displayed in Fig. 4. With five formula units for 
$\sqrt 5 \times \sqrt 5$ superlattice there can be two ions of Fe in
the unit cell with three possible distributions of Fe ions with respect to 
each other. Corresponding Fe ions are marked as large circles in Fig. 4.
Explicit LSDA calculations were made for the most symmetric case shown on the 
panel (a) of Fig. 4. The other two distributions of Fe ions produce the
smaller values of Curie temperature, as was shown for similar situation 
in Ref. \cite{Medvedev}.

This LSDA calculation for (Li$_4$Fe)(OH)$_5$(FeSe)$_5$ gives the ferromagnetic 
solution in agreement with LSDA calculation presented in Ref. \cite{Pachmayr}, 
with the value of magnetic moment on iron within Li$_4$Fe layer 
$\mu_{Fe}$ = 3.51$\mu_B$. Corresponding densities of states for 3d states of 
Fe ion within Li$_4$Fe(OH)$_5$ layer (solid line) and Fe-3d states in FeSe 
layer (dashed line) are plotted in Fig. 5.
We can see that 3d states of Fe ion within Li$_4$Fe(OH)$_5$ layer produce rather
narrow spikes in the density of states, which correspond to small amount of 
Fe ions in the layer. However, the introduction of Fe ions into LiOH layer 
increases the value of $N(E_F)$ up to 4.55 states/cell/eV, and in accordance 
with the estiamtes of Ref. \cite{PvsC}, this leads to the increase of $T_c$ 
up to 45K. Thus, the introduction of additional Fe ions into LiOH layers is 
important to obtain better agreement with experimental value of $T_c$=43K 
\cite{Pachmayr}.

\begin{figure}[!h]
\includegraphics[clip=true,width=0.5\textwidth]{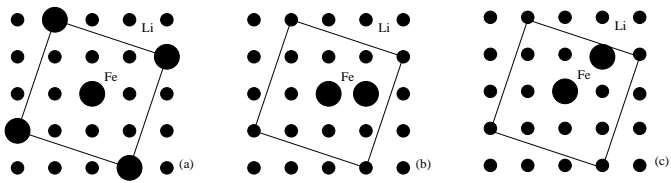}\\
\caption{Fig. 4. Possible distributions of Fe ions in Li$_4$Fe layer.
Large circles mark Fe ions, small -- Li. Solid line shows 
$\sqrt 5 \times \sqrt 5$ unit cell.} 
\end{figure}

To estimate the Curie temperature we have calculated the values of exchange 
parameters for the classical Heisenberg model at $T=0$, using the method 
proposed in Ref.~\cite{leip}. The value of nearest neighbor exchange integral 
obtained for Fe configuration, shown in on Fig. 4 (a) is $J$=1.3K.
Employing this calculated value of exchange parameter $J$ along with the
spin value $S$=2 (corresponding to experimental observation of Fe$^{2+}$
M\"ossbauer line in Ref. \cite{Pachmayr}) we obtain the Curie temperature 
$T_C=Jz\frac{S(S+1)}{3}$=10.4K (where $z$=4 is the number of nearest neighbors), 
which is quite close to experimental value of $T_C$=10K \cite{Pachmayr}.

\begin{figure}[!hb]
\includegraphics[clip=true,width=0.45\textwidth]{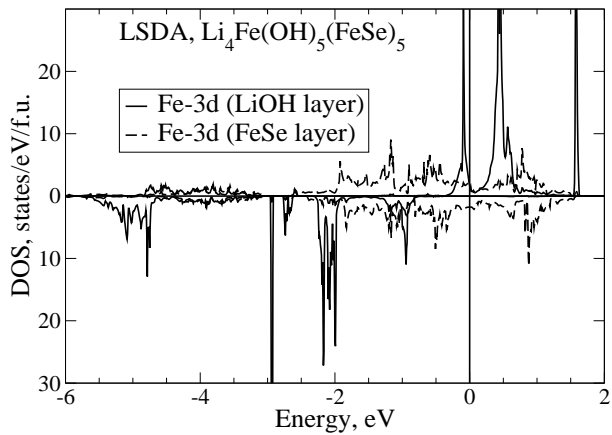}\\
\caption{Fig. 5. LSDA calculated densities of states of 
(Li$_4$Fe)(OH)$_5$(FeSe)$_5$.
Solid line corresponds to 3d states of Fe ion within Li$_4$Fe(OH)$_5$ layer,
dashed one --- Fe-3d states in the FeSe layer.
The Fermi level $E_F$ is at zero energy } 
\end{figure}

\section {Conclusion}

In this paper we have studied the paramagnetic band structure of the new 
iron -- based high-temperature superconductor
[Li$_{1-x}$Fe$_x$OH](Fe$_{1-y}$Li$_y$)Se (LiOHFeSe) with $T_c$=43K 
\cite{Pachmayr} by means of LDA band structure calculations, presenting the
band dispersions, densities of states and Fermi surfaces. These LDA calculated 
bands in the vicinity of the Fermi level as well as Fermi surfaces
are typical to other iron -- based superconductors. This is obviously due to
fact, that the FeSe layers of LiOHFeSe are isostructural to those in the bulk 
FeSe.

Doping the LiOH layer with Fe ions leads to ferromagnetic ordering of 
these ions in the layer. Our LSDA calculations confirmed the formation
of ferromagnetic structure in LiOH layers and calculated value of exchange
interaction has produced the theoretical value of Curie temperature 
$T_C$=10.4K, which agrees rather well with the experimental value of 10K 
\cite{Pachmayr}.

We thank M.V. Medvedev for useful discussions. This work is partly supported 
by RFBR grant No. 14-02-00065.


\begin{thebibliography}{99}


\bibitem{kamihara_08} Y. Kamihara, T. Watanabe, M. Hirano, H. Hosono. 
J. Am. Chem. Soc. {\bf 130}, 3296-3297 (2008).

\bibitem{UFN}M.V. Sadovskii, Uspekhi Fiz. Nauk {\bf 178}, 1243 (2008);
Physics Uspekhi {\bf 51}, No. 12 (2008).

\bibitem{Hoso_09}K. Ishida, Y. Nakai, H. Hosono. 
J.Phys. Soc. Jpn. {\bf 78}, 062001 (2009).

\bibitem{Johnson} D.C. Johnson, Advances in Physics {\bf 59}, 803 (2010).

\bibitem{FeSe}Y. Mizuguchi, Y. Takano. J. Phys. Soc. Jpn. {\bf 79}, 102001 (2010).

\bibitem{Mazin}P.J. Hirshfeld, M.M. Korshunov, I.I. Mazin. Rep. Prog. Phys.
{\bf 74}, 124508 (2011)

\bibitem{PvsC} M.V. Sadovskii, E.Z. Kuchinskii, I.A. Nekrasov,
JMMM {\bf 324} 3481, (2012).

\bibitem{relcomp} I.A. Nekrasov, M.V. Sadovskii,
Pis'ma Zh. Eksp. Teor. Fiz. {\bf 99}, 687 (2014) 
[JETP Letters, 99, No. 10 (2014)].


\bibitem{Pachmayr} U. Pachmayr, F. Nitsche, H. Luetkens, S. Kamusella, F. 
Br\"uckner, R. Sarkar, H.-H. Klauss, D, Johrendt, Angew. Chem. Int. Ed.
{\bf 53} (2014); DOI: 10.1002/anie.201407756; arXiv:1409.3982.

\bibitem{LMTO}O.K. Andersen. Phys. Rev. B {\bf 12}, 3060 (1975);
O. Gunnarsson, O. Jepsen,  O.K. Andersen. Phys. Rev. B {\bf 27}, 7144 (1983);
O.K. Andersen, O. Jepsen.  Phys. Rev. Lett. {\bf 53}, 2571  (1984).

\bibitem{Nekr2}I.A. Nekrasov, Z.V. Pchelkina, M.V. Sadovskii. Pis'ma Zh. Eksp.
Teor. Fiz. {\bf 88}, 155 (2008) [JETP Letters {\bf 88}, 144 (2008)]. 


\bibitem{SinghFeSe} A. Subedi, L. Zhang, D. J. Singh, and M. H. Du. 
Phys. Rev. B {\bf 78}, 134514 (2008).


\bibitem{Medvedev} M.V. Medvedev, I.A. Nekrasov,
Fizika Metallov i Metallovedenie [The Physics of Metals and Metallography],
to be published; arXiv:1008.3219.

\bibitem{leip}A.~I.~Liechtenstein, M.~I.~Katsnelson, V.~P.~Antropov, 
V.~A.~Gubanov, J. Magn. Magn. Mater. {\bf 67}, 65 (1987);
V.~I.~Anisimov, F.~Aryasetiawan, A.~I.~Lichtenstein, 
J.~Phys.: Condens. Matter {\bf 9}, 767 (1997).


\end{thebibliography}
\end{document}